\documentclass{ws-ijgmmp}
\usepackage{bigints}
\newcommand{\beqa}{\begin{eqnarray}}
\newcommand{\eeqa}{\end{eqnarray}}
\newcommand{\nn}{\nonumber}
\newcommand{\SL}{\mathrm{SL}(2,\mathbb R)}
\newcommand{\LL}{L^2(\mathrm{SL}(2,\mathbb R))}
\newcommand{\s}{\mathfrak{sl}(2,\mathbb R)}
\newcommand{\bpm}{\begin{pmatrix}}
\newcommand{\epm}{\end{pmatrix}}
\newcommand{\g}{\mathfrak{g}}
\def\d{\mathrm{d}}

\begin{document}

\markboth{Rutwig Campoamor-Strusberg, Alessio Marrani and Michel Rausch de Traubenberg}
{A Kac-Moody algebra associated to the
non-compact manifold SL$(2, \mathbb R)$}

%
\catchline{}{}{}{}{}
%

\title{A Kac-Moody algebra associated to the
non-compact manifold SL$(2, \mathbb R)$}

\author{Rutwig Campoamor-Stursberg}

\address{Instituto de Matem\'atica Interdisciplinar and Dpto. Geometr\'\i a y Topolog\'\i a, UCM, E-28040 Madrid, Spain\\
rutwig@ucm.es}

\author{Alessio Marrani}

\address{Instituto de F\'isica Te\'orica, Dep.to de F\'isica, Universidad de Murcia, Campus de Espinardo,
E-30100, Spain\\
alessio.marrani@um.es}

\author{Michel Rausch de Traubenberg}
\address{Universit\'e de Strasbourg, CNRS, IPHC UMR7178, F-67037 Strasbourg Cedex, France\\
Michel.Rausch@iphc.cnrs.fr}

\maketitle

\begin{history}
\received{(Day Month Year)}
\revised{(Day Month Year)}
\end{history}

\begin{abstract}
  We construct explicitly a Kac-Moody algebra associated to
  SL$(2, \mathbb R)$ in two different but equivalent ways: either by identifying a Hilbert basis of $L^2($SL$(2, \mathbb R))$ or by the Plancherel Theorem. Central extensions and Hermitean differential  operators  are identified.
\end{abstract}


\section{Introduction}	

Among infinite dimensional Lie algebras, affine Lie algebras are of utmost importance in theoretical and mathematical physics,  as shown {\it e.g.} by their applications in string theory,  $2D-$integrable
models, {\it etc.}). An affine Lie algebra, associated to
a Lie algebra $\g$, is a centrally extended infinite dimensional algebra associated to the Lie group $U(1)$ \cite{go}. Motivated by Kaluza-Klein theories,
extensions of affine Lie algebras associated to a compact Lie group $G_c$ or a coset $G_c/H$ were studied in \cite{rmm,rmm2}.
These algebras have indeed also been
considered by various authors in the case of the two-sphere \cite{bars}  or the $n-$tori \cite{bars}-\nocite{MRT, KT, Frap, RaSo}\cite{jap}.
The main tool to associate  a Lie algebra to a compact Lie group $G_c$ is to study $L^2(G_c)$,  the space of square integrable
functions on $G_c$. Moreover, the set of  matrix elements of all unitary
representations (once correctly normalised) is a orthonormal Hilbert
basis of $L^2(G_c)$ itself.

The purpose of this short note is to extend the construction
given in \cite{rmm} to the case of non-compact manifolds and groups, focusing on the simplest case, provided by the rank-1 Lie group $\SL$, and to partially reproduce the results
of \cite{ram}. Since the Lie group $\SL$ is non-compact, unitary representations of $\SL$ are infinite dimensional, and consequently the decomposition of functions in $\LL$ is much more involved.
In particular, irreducible unitary representations of $\SL$ decompose in two classes: the discrete
series and the continuous series \cite{bar}. The eigenvalues of the Casimir operator for the former
are discrete, whilst for the latter the eigenvalues are continuous. This implies that the matrix
elements of the discrete series
are normalisable whereas the matrix elements of the continuous series are non-normalisable. For such reasons, the harmonic analysis on $\SL$ is much more difficult than in the compact case. In general, the harmonic analysis on non-compact Lie groups is based on the Plancherel Theorem, which, in the case of $\SL$, yields that square integrable functions decompose as a sum over the discrete
series and an integral over continuous series. This decomposition is the main tool
to associate an infinite dimensional Lie algebra to a simple Lie algebra.
We will also present another equivalent construction of this infinite dimensional algebra, exploiting an explicit identification of a Hilbert basis of $\LL$.
It is finally shown that algebras  associated to $\SL$ admit central extensions.

\section{Representations of SL$(2,\mathbb R)$}\label{sec:rep}
In this section we briefly review  results concerning the Lie group SL$(2,\mathbb R)$, which are useful for the definition of the Kac-Moody algebra associated to the manifold SL$(2,\mathbb R)$. For further details, we address the reader to \cite{ram}.

\subsection{Irreducible unitary representations}\label{sec:irep}
Irreducible unitary representations of SL$(2,\mathbb R)$ were classified by Bargmann \cite{bar} (see also \cite{vk,ggv}).
Let $K_\pm, K_0$ be the generators of the Lie algebra $\mathfrak{sl}(2,\mathbb R)$ satisfying
\beqa
\label{eq:LB}
\big[K_0,K_\pm\big]=\pm K_\pm\ , \ \
\big[K_+,K_-\big]=-2 K_0\
\eeqa
and let $Q$ be the Casimir operator
\beqa
Q=K_0^2 -\frac 12\big(K_+ K_-+ K_- K_+\big) \ . \nn
\eeqa

Non-trivial unitary representations are divided in two classes. The discrete series, which are either bounded from
below ${\cal D}^+_\lambda=\{\big|\; \lambda,+,n\big>, n\ge \lambda>0\}$
  \beqa
  \label{eq:D+}
K_{0} \big|\lambda,+,n\big>&=&n\big|\lambda,+,n\big>\ ,  \notag   \\
K_{\pm}\big|\lambda,+,n\big> &=&\sqrt{(n\pm\lambda)(n\pm(1-\lambda))}\big|\lambda,+,n\pm1\big>\ ,  \\
Q\big|\lambda,+,n\big> &=&\lambda \left( \lambda -1\right) \big|\lambda,+,n\big>\ ,\notag
\eeqa
or bounded from above ${\cal D}^-_\lambda=\{\big|\lambda,-,n\big>, n\le- \lambda <  0\} $
 \beqa
   \label{eq:D-}
K_{0} \big|\lambda,-,n\big>&=&n\big|\lambda,-,n\big>;  \notag   \\
K_{\pm}\big|\lambda,-,n\big> &=&-\sqrt{(-n\mp\lambda)(-n\mp(1-\lambda))} \big|\lambda,-,n\pm 1\big> \\
 Q\big|\lambda,-,n\big> &=&\lambda \left( \lambda -1\right) \big|\lambda,-,n\big>\ .\notag
\eeqa
The representations ${\cal D}^\pm_\lambda$ are unitary if $\lambda \in \mathbb N\setminus\{0\}$ (bosonic representations)
or
$\lambda \in \mathbb N + 1/2$ (fermionic representations). Note that, in both cases, the eigenvalues of the Casimir operator are discrete.

The
continuous series  consists of two types: principal and supplementary.   The principal continuous series is  ${\cal C}^{i \sigma,\epsilon}=\{\big|\sigma, \epsilon,n\big>, n \in \mathbb Z + \epsilon\}$ with $\epsilon=0$ (resp. $\epsilon=1/2$) for bosonic  (resp. fermionic) representations, with
\beqa
\label{eq:Cont}
K_0 \big|\sigma, \epsilon,n\big>&=& n\big|\sigma, \epsilon,n\big>\ , \nn\\
K_\pm \big|\sigma, \epsilon,n\big>&=& \sqrt{(n\pm\frac 12+\frac i2 \sigma)(n\pm\frac12-\frac i2\sigma)}\big|\sigma, \epsilon,n\pm1\big>\ , \\
Q\big|\sigma, \epsilon,n\big>&=& \left(\frac 12 +i\sigma\right)\left(\frac 12 +i\sigma -1\right)\big|\sigma, \epsilon,n\big>= -\left(\frac 14 +\sigma^2\right)\big|\sigma, \epsilon,n\big> \ . \nn
\eeqa
The representation is unitary if $\sigma>0$, and the eigenvalues of the Casimir operator  (denoted $q$) are continuous and upper bounded by $q<-1/4$.  The supplementary continuous series is  ${\cal C}^{ \sigma}=\{\big|\sigma,n\big>, n \in \mathbb Z \}$ (admitting bosonic representations only), with
\beqa
K_0 \big|\sigma,n\big>&=& n\big|\sigma,n\big>\ , \nn\\
K_\pm \big|\sigma,n\big>&=& \sqrt{(n\pm\frac 12+\frac12 \sigma)(n\pm\frac12-\frac12\sigma)}\big|\sigma,n\pm1\big>\ , \nn \\
Q\big|\sigma,n\big>&=& \left(\frac 12 +\sigma\right)\left(\frac 12 +\sigma -1\right)\big|\sigma,n\big>=\left(-\frac 14 +\sigma^2\right)\big|\sigma,n\big> \ . \nn
\eeqa
The unitary condition $0< \sigma^2 <1/4$ is equivalent to the condition $-1/4<q<0$.

\subsection{Matrix elements}\label{sec:mat}
A group isomorphism characterising SL$(2,\mathbb R)$ is $\SL \cong SU(1,1)$, thus it can be defined by the set of $2\times 2$ complex matrices
\beqa
SU(1,1) = \Bigg\{U = \begin{pmatrix} z_1 & z_2 \\
 \bar z_2   & \bar z_1 \end{pmatrix}  \ , \ \ z_1, z_2 \in \mathbb C:  \ \ |z_1|^2-|z_2|^2 =1 \Bigg\} \ . \nn
\eeqa
A convenient parameterisation reads
\begin{equation}
\label{eq:param}
z_{1}=\cosh \rho e^{i\varphi _{1}}\ ,\ \ z_{2}=\sinh \rho e^{i\varphi _{2}}\
,\ \ \rho \geq 0\ ,0\leq \varphi _{1},\varphi _{2}<2\pi
\end{equation}%
with the scalar product of functions defined as
\begin{equation}
\label{eq:sp}
(f,g)=\frac{1}{ 4\pi^{2}}\int\limits_{0}^{+\infty }\cosh \rho \sinh
\rho \text{d}\rho \int\limits_{0}^{2\pi }\text{d}\varphi
_{1}\int\limits_{0}^{2\pi }\text{d}\varphi _{2}\bar{f}(\rho ,\varphi
_{1},\varphi _{2})g(\rho ,\varphi _{1},\varphi _{2})\ .
\end{equation}%
Within the parameterisation \eqref{eq:param},
 the generators of $\s$ for  the left action (denoted by $L$) take the form
\begin{eqnarray}
\label{eq:L}
L_{\pm} =\frac{1}{2}e^{i(\varphi _{1}\mp\varphi _{2})}\Big[i\tanh \rho
\;\partial _{1}\pm\partial _{\rho }-i\coth \rho \;\partial _{2}\Big]\ , \ \
L_{0} =\frac{i}{2}\big(\partial _{2}-\partial _{1})\ ,
\nn
\end{eqnarray}%
whereas the ones for  the right action (denoted by $R$) read
\begin{eqnarray}
\label{eq:R}
R_{\pm} =\frac{1}{2}e^{\pm i(\varphi _{1}+\varphi _{2})}\Big[-i\tanh
\rho \;\partial _{1}\mp\partial _{\rho }-i\coth \rho \;\partial _{2}\Big]\ , \ \
R_{0} =-\frac{i}{2}\big(\partial _{2}+\partial _{1})\ .
\nn
\end{eqnarray}%
The Casimir operator reads
\beqa
Q &=&\frac{1}{2}\coth(2 \rho)\partial_{\rho }+\frac{1}{4}\partial _{\rho }^{2}-\frac{1-\tanh^{2}\rho}{4}\partial _{1}^{2}+\frac{\coth^2\rho-1}{4}\partial _{2}^{2} \ .
\notag
\eeqa

\textit{\c{C}a va sans dire} that the generators for the left and right action satisfy the commutations relations \eqref{eq:LB}, and that the $L$'s commute with the $R$'s.

The matrix elements of the irreducible unitary representations of SL$(2,\mathbb R)$ were obtained by Bargmann in \cite{bar}.
We now review the main steps to obtain all matrix elements, but with a slightly different method  than that used by Bargmann.
Let  $\Psi_{n,\Lambda,m}$ be the matrix elements of the representations given in Sec.  \ref{sec:irep},
with $\Lambda=(\lambda,+)$ or $(\lambda,-)$ for the discrete series and $\Lambda=(i\sigma,\epsilon)$ (with $\epsilon=0,1/2$) for the
principal continuous series (since the supplementary series plays no r\^ole, we will henceforth ignore  it).
\begin{enumerate}
\item We first solve the differential equations
\beqa
\label{eq:SolQ}
L_0\Psi_{n,\Lambda,m}(\rho, \varphi_1, \varphi_2)&=& n \Psi_{n,\Lambda,m}(\rho, \varphi_1, \varphi_2)\ ,\nn\\
R_0 \Psi_{n,\Lambda,m}(\rho, \varphi_1, \varphi_2)&=& m \Psi_{n,\Lambda,m}(\rho, \varphi_1, \varphi_2)\ ,\\
Q  \Psi_{n,\Lambda,m}(\rho, \varphi_1, \varphi_2)&=& q \Psi_{n,\Lambda,m}(\rho, \varphi_1, \varphi_2 ) \ ,\nn
\eeqa
where the possible values of $m,n,q$ are given in Sec.  \ref{sec:irep} and depend on the representation.
The first two equations of \eqref{eq:SolQ} are solved by
\beqa
\Psi_{n,\Lambda,m}(\rho, \varphi_1, \varphi_2)=e^{i(n+m)\varphi_1 + i (m-n)\varphi_2} f_{n,\Lambda,m}(\rho) \ ,\nn
\eeqa
and the third equation enables us to express $f_{m,\Lambda,n}$ in term of hypergeometric functions
(or hypergeometric polynomials for the discrete series). The functions
$\Psi_{n,\Lambda,m}$ can then be determined up to a coefficient $C_{n,q,m}$.
\item We now impose that the operators $L_\pm$ and $R_\pm$ have the right action
  on $\Psi_{n,\Lambda,m}$ (see \eqref{eq:D+},
  \eqref{eq:D-} and \eqref{eq:Cont}).  The functions
$\Psi_{n,\Lambda,m}$ are then given up to a coefficient $C_{q}$.
\item For the discrete series, we impose the normalisation $\|\Psi_{n,\lambda,\pm,m }\|=1$
with respect to the scalar product \eqref{eq:sp}, whereas
  for the continuous series we set $\Psi_{n,i\sigma,\epsilon,m}(0,0,0)=\delta_{mn}$. This fixes completely $\Psi_{n,q,m}$
  (precisely, up to a phase -- that we take equal to 1 -- for the discrete series).
\end{enumerate}
In any case, the expression for $\Psi_{n,\Lambda,n}$ depends on whether $n\ge m$ or $m \ge n$. We  restrict our analysis to some of the resulting expressions for the functions $\Psi_{n,\Lambda,n}$. For instance, for
the discrete series bounded from below and for $m \ge n$, one obtains
\beqa
&\Psi_{n,\lambda,+,n}(\rho,\varphi_1,\varphi_2)=
  \frac {\sqrt{ 2(2\lambda-1)}} {(m-n)!} \sqrt{\frac{(m-\lambda)!(m+\lambda-1)!}
    {(n-\lambda)!(n+\lambda-1)!}} e^{i(m+n)\varphi_1 + i(m-n)\varphi_2}
  \times \nn\\
&   \cosh^{-m-n}\rho \sinh^{m-n}\rho \;  {}_2F_1(-n+\lambda,-n-\lambda+1;1+m-n;-\sinh^2\rho),  \nn
  \eeqa
  with $\lambda \in \mathbb N \setminus \{0\}$ or
 $\lambda \in \mathbb N \setminus \{0\}+ 1/2$  and $m,n\ge \lambda$.
      On the other hand, for the principal continuous series and for $m\ge n$,  we get
\beqa
&\Psi_{n,i \sigma,\epsilon,n}(\rho,\varphi_1,\varphi_2)=
\frac 1 {(m-n)!} \sqrt{\frac{\Gamma(m+\frac 12 +\frac 12 i\sigma)\Gamma(m+\frac 12 -\frac 12 i\sigma)}
                              {\Gamma(n+\frac 12 +\frac 12 i\sigma)\Gamma(n+\frac 12 -\frac 12 i\sigma)}}
  e^{i(m+n)\varphi_1 +i(m-n)\varphi_2} \times\nn\\
  &\cosh^{m+n}\rho \sinh^{m-n}\rho \; {}_2F_1(m+\frac12 + \frac12 i\sigma,m+\frac12 - \frac12 i\sigma;m-n+1; -\sinh^2 \rho),\nn
  \eeqa
  with $\sigma>0$ and $m,n\in \mathbb Z + \epsilon$.  In both cases ${}_2 F_1(\alpha,\beta;\gamma;z)$ is a hypergeometric function,
  which reduces to a hypergeometric polynomial for the discrete series.

  Note that the set of matrix elements of the discrete series bounded from below and from above constitutes an orthonormal set with respect to the scalar product
  \eqref{eq:sp}:
  \beqa
  \label{eq:orth}
  (\Psi_{n,\lambda,\eta,m}, \Psi_{n',\lambda',\eta',m'})=\delta_{nn'} \delta_{mm'}
  \delta_{\lambda \lambda'} \delta_{\eta \eta'},
  \eeqa
  with $\eta,\eta'= \pm$.
  We should also remark that the representation corresponding to $\lambda=1/2$ is not normalisable, implying that $\lambda>1/2$, whereas none of the matrix elements of the principal continuous series is normalisable, since
  \beqa
  \label{eq:nnor}
(\psi_{n,i \sigma,\epsilon, m}, \psi_{n',i \sigma',\epsilon', m'})=\frac{1}{\sigma\tanh\pi(\sigma+i\epsilon)} \delta_{\epsilon \epsilon'}
\delta_{mm'} \delta_{n n'} \delta(\sigma-\sigma')\ .
\eeqa

\section{Square integrable functions $\LL$}
Our construction of the Kac-Moody Lie algebra associated to $\SL$ strongly relies on the set of square integrable functions $\LL$. We will now summarise the main results which enable us to expand square integrable functions of $\LL$ by exploiting the Plancherel Theorem, or to identify a Hilbert basis of $\LL$.

\subsection{Plancherel Theorem} \label{sec:plan}
Given a compact Lie group $G_c$, the Peter-Weyl Theorem \cite{PW} states that the matrix elements of all its unitary
representations (once correctly normalised) constitute an orthonormal Hilbert
basis of the set $L^2(G_c)$ of square integrable functions on $G_c$. The
 situation is much more complicated for a non-compact Lie group, like $\SL$. Indeed, in Sec. \ref{sec:mat} we have seen that the matrix
elements of the discrete series (bounded from below and above) are an orthonormal set (see \eqref{eq:orth}), but this set is
not complete and thus is not a Hilbert basis of $\LL$. On the other hand, the
matrix elements of the principal continuous series are not
normalisable (see \eqref{eq:nnor}), and consequently, they do not belong to
$\LL$. In the non-compact case, one has indeed to resort to the Plancherel Theorem \cite{hc,sch}, which enables the expansion of the square integrable functions of $\SL$ in terms of a sum over the discrete series and of an
integral over the principal continuous series.

Let us now introduce a Gel'fand triple of $\SL$, defined by ${\cal S}, {\cal S}'$ such that
\beqa
{\cal S} \subset \LL \subset {\cal S}' \ , \nn
\eeqa
where ${\cal S}$ denotes the space of functions which decrease rapidly in the $\rho-$ direction (see {\it e.g.}  \cite{Va} Chap. 8); they are also called Schwartz functions, and form a dense subspace of $\LL$. On the other hand, ${\cal S}'$ is the dual space of ${\cal S}$. Remarkably, the matrix elements of the continuous principal series
$\Psi_{n i\sigma \epsilon  m}$ belong to ${\cal S}'$ (see for instance \cite{Ba-Ra}).
Moreover, given a function $f$ in ${\cal S}$, it holds that \cite{Ba-Ra,Va}
\beqa
\label{eq:Plan}
f(\rho,\varphi_1,\varphi_2) &=&\sum \limits_{\eta=\pm}\sum \limits_{\lambda>\frac12}\;
\sum \limits_{\eta m,\eta n\ge \lambda}\; f^{n\lambda \eta m}\; \Psi_{n\lambda \eta  m}(\rho,\varphi_1,\varphi_2)\\
&&  +\sum \limits_{\epsilon = 0, 1/2}\sum \limits_{m,n \in \mathbb Z + \epsilon}\int \limits_0^{+\infty} \text{d} \sigma\; \sigma \tanh (\pi \sigma +i \epsilon)
 f^{n   \epsilon m}(\sigma) \psi_{n i \sigma \epsilon m}(\rho,\varphi_1,\varphi_2)\nn
 \eeqa
 with
 \beqa
 \label{eq:PC}
 f^{n \lambda \eta m}&=&(\Psi_{n\lambda  \eta m},f) \\
 f^{n \epsilon m}(\sigma)&=& (\psi_{ni\sigma \epsilon  m},f) \ . \nn
 \eeqa
The r.h.s. of \eqref{eq:Plan} can be rewritten in the following concise way:
\beqa
\label{eq:gene}
f(\rho,\varphi_1,\varphi_2) &=&   \sum_{\Lambda, n, m} \hskip -.55 truecm \int{}\;
f^{n \Lambda m}
\Psi_{n\Lambda m }(\rho,\varphi_1,\varphi_2)
\eeqa
where $\Lambda= (\lambda,+), (\lambda,-), (i\sigma,0)$ or $(i\sigma,1/2)$, and the symbol $\sum\hskip -.35 truecm \int{}\; $ indicates a summation over the discrete values of $\Lambda$ and an integration over its continuous values.
 \subsection{Hilbert basis}\label{sec:hil}
As given by \eqref{eq:Plan}, the Plancherel Theorem gives rise to an expansion of the Schwartz functions as a sum over the matrix elements of the discrete series and as an integral over the matrix elements of the continuous series. However since the former do not constitute a complete set and the latter are not normalisable, the whole set of matrix elements does not constitute a Hilbert basis of $\LL$. However, it is well known (see {\it e.g.} \cite{RS, GP}) that any Hilbert space admits a Hilbert basis, {\it i.e.}, a complete
 countable set of orthonormal vectors. Viktor Losert (Institut f\"ur Mathematik, Universit\"at Wien, viktor.losert@univie.ac.at) identified
 for us a Hilbert basis of $\LL$; below, we recall his results (again, for more details, please see \cite{ram}).

We introduce the eigenspaces of the operators $L_0$ and $R_0$
\beqa
W_{nm} = \Bigg\{ F \in \LL\ , \ \ F(\rho,\varphi_1,\varphi_2)=e^{i(m+n)\varphi_1 + i(m-n)\varphi_2} f(\rho) \Bigg\} \ ,\nn
\eeqa
such that, for any $F\in W_{nm}$, it holds that
\beqa
L_0 F(\rho,\varphi_1,\varphi_2)&=&n F(\rho,\varphi_1,\varphi_2) \ , \nn\\
R_0 F(\rho,\varphi_1,\varphi_2)&=&m F(\rho,\varphi_1,\varphi_2) \ . \nn
\eeqa
A Hilbert basis of  $W_{mn}$ is provided by
\beqa
    {\cal B}_{nm} = \Bigg\{\Phi_{nnk}(\rho,\varphi_1,\varphi_2) = e^{i(m+n)\varphi_1 +i(m-n)\varphi_2} e_{nmk}(\cosh 2 \rho), k \in \mathbb N \Bigg\}, \nn
    \eeqa
where the functions $e_{nmk}(\cosh 2 \rho)$ have to be determined. In general, $\LL$ can be split as follows:
    \beqa
\LL= \LL^d \oplus \LL^{d^\perp}, \nn
\eeqa
where $\LL^d$ denotes the set of square integrable functions whose Hilbert basis is provided by the matrix elements of the discrete series (bounded from below and above), while $\LL^{d^\perp}$ stands for the set of square integrable functions orthogonal to $\LL^d$.
    Three cases have to be considered  for $W_{nm}$ \cite{ram}:
    \begin{enumerate}
    \item $mn >0, m,n>1/2$:  $ {\cal B}_{nm}$ contains matrix elements of the discrete series bounded from below, together with elements
      of $\LL^{d^\perp}$.
   \item $mn >0, m,n<-1/2$: $ {\cal B}_{nm}$ contains matrix elements of the discrete series bounded from above, together with elements
     of $\LL^{d^\perp}$.
     \item $mn<0$ or $m=0,1/2$ or $n=0,1/2$: $ {\cal B}_{nm}$ contains only elements
     of $\LL^{d^\perp}$.
    \end{enumerate}
   Here, we will only present the explicit expression of Hilbert basis of $W_{mn}$ pertaining to the case (1) above (see \cite{ram} for the other cases). As in Sec \ref{sec:mat},
    the functions $e_{nmk}$ depend on whether $n\ge m$ or $m \ge n$. For instance, for $n\ge m$ (and defining $x:=\cosh 2 \rho$) the elements of $\LL^d$ read
     \beqa
\label{eq:ed}
e_{nmk}(x)&=& \sqrt{2^{2m-1}\frac{(2m-2k-1)k!(m+n-k-1)!}{(2m-k-1)!(n-m+k)!} } \times\\
&&(x-1)^\frac{n-m}2 (x+1)^{-\frac{m+n}2} P_k^{(n-m,-n-m)}(x) \ , 0\le k< m-\epsilon,\nn
\eeqa
whereas the elements of $\LL^{d^\perp}$ read
    \beqa
     \label{eq:e}
e_{nmk}(x) =  \sqrt{2^{2k + 2 \epsilon+1}\frac{(2k +n-m + 2\epsilon+1) (k +n-m + 2\epsilon) !k! }{(k+2\epsilon)! (n-m+k)!} } \times\nn\\
    (x-1)^{\frac{n-m}2} (x+1)^{\frac{m-n}2 -k-\epsilon -1}
    P_k^{(n-m,m-n-2k-2\epsilon-1)}(x) \ , k\ge m -\epsilon,
    \eeqa
    where $\epsilon=0$ if $m,n$ integer, and $\epsilon=1/2$ if $m,n$ half-integer. In \eqref{eq:ed} and \eqref{eq:e}, $P^{(a,b)}_k$ denotes the Jacobi polynomials (it is here worth observing that, for the matrix elements of the discrete series, it is possible to express the hypergeometric function given in \eqref{eq:D+} and \eqref{eq:D-} in terms of the Jacobi polynomials themselves \cite{ram}).
    
    Some observations are in order. In the case (1) above
    and for $n\ge m$, when
    $k<m-\epsilon$ then $\Phi_{nmk} \in W_{nm}\cap \LL^d$,  and when $k\ge m-\epsilon$ then  $\Phi_{nmk} \in W_{nm}\cap \LL^{d^\perp}$. This is a general
    property of  $ {\cal B}_{nm}$ : indeed, in the cases (1) and (2) above, when $k\ge k_{\text{min}}\ne 0$
    then $\Phi_{nmk} \in W_{nm}\cap \LL^{d^\perp}$, and when  $0\le k<k_{\text{min}}$ then
    $\Phi_{nmk} \in W_{nm}\cap \LL^{d}$.
    On the other hand, in the case (3) above, when $k\ge k_{\text{min}}= 0$  then $\Phi_{nmk} \in W_{nm}\cap \LL^{d^\perp}$; this can be traced back to the fact that, in this case, $W_{nm}\cap \LL^d = \emptyset$. In \cite{ram} the precise value of $k_{\text{min}}$ in all cases has been computed.
    
     It follows that the set $\cup_{n,m \in \mathbb Z} {\cal B}_{nm} \cup_{n,m \in \mathbb Z+1/2} {\cal B}_{nm}$ is an orthonormal Hilbert basis of $\LL$:
    \beqa
(\Phi_{nmk},\Phi_{n'm'k'})= \delta_{nn'} \delta_{mm'} \delta_{kk'} \ . \nn
    \eeqa
    Therefore, for any
    $f \in \LL$, it holds that
    \beqa
    \label{eq:los}
    f(\rho,\varphi_1,\varphi_2) &=& \sum \limits_{\epsilon=0,1/2} \sum \limits_{n,m \in \mathbb Z +\epsilon}
    \sum \limits_{k=0}^{+\infty} f^{nmk} \Phi_{nmk}(\rho,\varphi_1,\varphi_2)\ , \\
    f^{nmk}&=&(\Phi_{mnk},f) \ . \nn
    \eeqa

    Since the   $\mathfrak{sl}(2,\mathbb R)$ generators act on $\LL$, and since $\LL^d \subset \LL$  is a sub-representation, {\it i.e.}, an invariant subspace of $\LL$, then  $\LL^{d^\perp}$ is also a representation of $\SL$. Indeed, for $m\ge n$ the action of $L_\pm, R_\pm$ on $\Phi_{nmk} \in \LL^{d^\perp}$ is given by
  \beqa
  L_+\Phi_{nmk}(\rho,\varphi_1,\varphi_2) &=&\alpha^L_{nmk} \Phi_{ n+1m k+1}(\rho,\varphi_1,\varphi_2) + \beta^L_{nmk}  \Phi_{n+1mk}(\rho,\varphi_1,\varphi_2),\nn\\
  L_-\Phi_{nmk}(\rho,\varphi_1,\varphi_2) &=&\gamma^L_{nmk}  \Phi_{ n-1 nk-1} (\rho,\varphi_1,\varphi_2)+ \delta^L_{nmk}  \delta_{n-1mk}(\rho,\varphi_1,\varphi_2),\nn\\
  R_+\Phi_{nmk} (\rho,\varphi_1,\varphi_2)&=&\alpha^R_{nmk}  \Phi_{ nm +1k-1} (\rho,\varphi_1,\varphi_2)+ \beta^R_{nmk}  \Phi_{nm+1k}(\rho,\varphi_1,\varphi_2),\nn\\
  R_-\Phi_{nmk}(\rho,\varphi_1,\varphi_2) &=&\gamma^R_{nmk}  \Phi_{nm-1 k+1}(\rho,\varphi_1,\varphi_2) + \delta^R _{nmk}  \Phi_{nm-1k}(\rho,\varphi_1,\varphi_2),\nn
  \eeqa
  with similar expressions for $n\ge m$. Furthermore, the action of the Casimir operator reads
  \beqa
  &Q\Phi_{nmk}(\rho,\varphi_1,\varphi_2) = a_{nmk} \Phi_{nmk-1}(\rho,\varphi_1,\varphi_2)\nn\\
  &+ b_{nmk} \Phi_{nmk}(\rho,\varphi_1,\varphi_2)+
  c_{nmk} \Phi_{nmk+1}(\rho,\varphi_1,\varphi_2)\ . \nn
  \eeqa
  The above formulas, whose coefficients have been computed in \cite{ram}, clearly shows that the $\SL$-representation given by $\LL^{d^\perp}$ is unitary but {\it is not irreducible}.

   For what concerns the asymptotic ($\rho \rightarrow \infty $) behaviour of the functions $e_{nmk} \in W_{nm} \cap \LL$, the behaviour of $e_{nmk} \in W_{nm} \cap \LL^d$   was studied in \cite{bar}, and the analysis has then been extended for $e_{nmk} \in W_{nm} \cap \LL^{d^\perp}$ \cite{ram}. Remarkably, the functions $\Phi_{nmk}$ are Schwartz functions, implying in particular that for $\Phi_{nmk} \in W_{nm} \cap \LL^{d^\perp}$ the Plancherel Theorem yields that
\beqa
\label{eq:PL}
\Phi_{nmk}(\rho,\varphi_1,\varphi_2) = \int \limits_0^{+\infty} \text{d} \sigma \; \sigma\tanh \pi(\sigma+i\epsilon) f^{nmk}(\sigma) \Psi_{n i\sigma \epsilon m}(\rho)\ ,
\eeqa
where $\epsilon=0$ if $n,m$ are integer and $\epsilon=1/2$ if $n,m$ are half-integer.
Furthermore, by using \eqref{eq:PC} and recalling the definition of the scalar product \eqref{eq:sp}, it holds that
\beqa
f^{nmk}(\sigma)= (\Psi_{n i\sigma \epsilon m},e_{nmk}) \ ,\nn
\eeqa
and the formula \eqref{eq:PL} can be inverted as follows:
\beqa
\label{eq:LP}
\Psi_{n i\sigma \epsilon m}(\rho,\varphi_1,\varphi_2) = \sum \limits_{k\ge k_{\text{min}}} \overline{ f^{mnk}}(\sigma)\Phi_{nmk}(\rho,\varphi_1,\varphi_2) \ ,
\eeqa
where $k_{\text{min}}$ has been introduced above.
We are thus able to express the matrix elements of the  (principal) continuous series in terms of the Hilbert basis provided by V. Losert (which will henceforth be named \textit{Losert basis}), and conversely. As we will see, this fact will be important for the construction of the Kac-Moody Lie algebra associated to $\SL$.

 \subsection{Clebsch-Gordan coefficients}
 The Clebsch-Gordan coefficients corresponding to the coupling of two representations ${\cal D}_\Lambda\otimes {\cal D}_{\Lambda'}$ were studied in \cite{hb1, hb2}. The coupling of two discrete series was studied by means of a bosonic realisation of the Lie algebra $\s$, whilst when at least one continuous series is involved the result was obtained by an analytic continuation. We do not consider all the cases in this short note, but we rather only report two examples (again, see \cite{ram} for more details), namely :
 \begin{enumerate}
\item product of two discrete series bounded from below or above
  ($a=(\rho,\varphi_1,\varphi_2)$):
  \beqa
  \Psi_{m_1\lambda_1\pm m'_1}(a)\Psi_{m_2\lambda_2\pm m'_2}(a) =
\sum \limits_{\lambda \ge \lambda_1 + \lambda_2}
  C_{\pm \lambda_1, \pm \lambda_2}^{\pm \lambda }{}_{m_1,m_2,m_1',m_2'}
  \Psi_{m_1+m_2 \lambda \pm m'_1+m'_2}(a),\nn
  \eeqa
  with
\beqa
& C_{\pm \lambda_1, \pm \lambda_2}^{\pm \lambda }{}_{m_1,m_2,m_1',m_2'}=
\frac{(4 \lambda_1-2)(4 \lambda_2-2)}{4 \lambda-2}
\scriptsize
\bpm\pm;\lambda_1&\pm;\lambda_2&\pm;\lambda\\
m_1&m_2&m_1+m_2\epm
\overline{\bpm\pm;\lambda_1&\pm;\lambda_2&\pm;\lambda\\
m'_1&m'_2&m'_1+m'_2\epm}.
\nn
\eeqa
\item  product of  one  discrete series bounded from below and one discrete series bounded from  above:
  \beqa
  &\Psi_{m_1\lambda_1+m'_1}(a) \Psi_{m_2\lambda_2-m'_2}(a)=\hskip -.7truecm
  \sum\limits_{\frac12<\lambda\le |\lambda_1 -\lambda_2|}  \hskip -.4truecm  C_{+\lambda_1,- \lambda_2}^{\eta_{12} \lambda}{}_{m_1,m_2,m_1',m_2'}
  \Psi_{m_1+m_2 \lambda\eta_{12} m_1'+m'_2}(a)  \nn\\
  &+ \int\limits_0^{+\infty}\text{d} \sigma \; \sigma \tanh \pi(\sigma+i\epsilon_{12}) C_{+\lambda_1,- \lambda_2}^{\epsilon_{12} i\sigma}{}_{m_1,m_2,m_1',m_2'}
  \Psi^{\epsilon_{12} }_{m_1+m_2 i\sigma m'_1+m'_2}(a)\nn
\eeqa
(with $\eta_{12}$ denoting the sign of $\lambda_1-\lambda_2$, and $\epsilon_{12}=0,1/2$ depending if the r.h.s. of the formula above is a boson or a fermion), where
\beqa
 C_{+\lambda_1,- \lambda_2}^{\eta_{12} \lambda}{}_{m_1,m_2,m_1',m_2'}&=&\frac{(4 \lambda_1-2)(4 \lambda_2-2)}{4 \lambda-2}\scriptsize
\bpm +;\lambda_1&-;\lambda_2&\eta_{12};\lambda\\
 m_1&m_2&m_1+m_2\epm
 \overline{\bpm +;\lambda_1&-;\lambda_2&\eta_{12};\lambda\\
 m'_1&m'_2&m'_1+m'_2\epm},\nn\\
 \normalsize
C_{+\lambda_1,- \lambda_2}^{\epsilon_{12} i\sigma}{}_{m_1,m_2,m_1',m_2'} &=&
 \scriptsize
 \bpm +;\lambda_1&-;\lambda_2&\epsilon_{12};i\sigma\\
 m_1&m_2&m_1+m_2\epm
 \overline{ \bpm +;\lambda_1&-;\lambda_2&\epsilon_{12};i\sigma\\
 m'_1&m'_2&m'_1+m'_2\epm}\ . \nn
\eeqa
 \end{enumerate}
In the above formulas, {\tiny  $\bpm \ \ \cdots \ \ \\ \cdots\epm$} denotes the corresponding Clebsch-Gordan coefficients.\\

In general, the product of two generic matrix elements involves both a sum over discrete series and an integral over the continuous series. Within the notation introduced in \eqref{eq:gene}, this fact is expressed as
\beqa
\label{eq:matmat}
\Psi_{m_1\Lambda_1  m'_1}(a) \Psi_{m_2\Lambda_ 2m'_2}(a) &=&   \sum_{\Lambda} \hskip -.5 truecm \int{}\;
{ C}_{\Lambda_1, \Lambda_2}^\Lambda{}_{m_1,m_2, m'_1,m'_2}
\Psi_{m_1+m_2 \Lambda m'_1+m'_2}(a)
\eeqa
where $\Lambda_1, \Lambda_2= (\lambda,+), (\lambda,-), (i\sigma,0)$ or $(i\sigma,1/2)$ and  $\Lambda$ takes one of the allowed  values
occurring in tensor product decomposition.\\

Similarly, if one uses the Losert basis, since the product $e_{nmk}(x) e_{n'm'k'}(x)$ is square integrable (and, even better, is a Schwartz function \cite{ram}), it holds that
\beqa
\label{eq:ee}
\Phi_{nmk}(a) \Phi_{n'm'k'}(a) = \sum \limits_{k''} C^{k''}_{kk'nn'mm'} \Phi_{n+n'm+m'k''}(a)
\eeqa

\section{The Kac-Moody algebra $\widehat{\g}(\SL)$}

In  \cite{rmm, rmm2} Kac-Moody algebras associated  to a  compact Lie group $G_c$ or to a coset  space $G_c/H$ were defined, by means of the Peter-Weyl Theorem. In this section, we will briefly review the results of \cite{ram}, in which a Kac-Moody algebra associated  to the non-compact Lie group $\SL$ was introduced.
\subsection{Construction}
As anticipated above, the Plancherel Theorem and the identification of a  Hilbert basis of $\LL$ are central in this construction, which can be split into the following steps. \\

(1): Let $\g$ be a simple (complex or real) Lie algebra, with basis $\{T^a, a=1,\cdots,\dim \g\}$, Lie brackets
\beqa
\big[T^a, T^b\big] = i f^{ab}{}_c T^c \ , \nn
\eeqa
and Killing form
\beqa
\Big<T^a, T^b\Big>_0=g^{ab} = \text{Tr}\Big(\text{ad}(T^a)\; \text{ad}(T^b)\Big) \ . \nn
\eeqa

(2): Let $\g(\SL)$ be the space of smooth maps from $\g$ to $\SL$. We can then expand any element $T^a(\rho,\varphi_1,\varphi_2)$ $\in$  $\g(\SL)$ by using either the Plancherel Theorem (discussed in Sec. \ref{sec:plan}) or the Hilbert basis of $\LL$ (discussed in Sec. \ref{sec:hil}). To streamline the presentation, the expansion based on \eqref{eq:Plan} will be named the \textit{Plancherel basis} (PB), and we have already named the expansion based on \eqref{eq:los} as the \textit{Losert basis} (LB). Thus,
\beqa
T^a(\rho,\varphi_1,\varphi_2)= \left\{
\begin{array}{ll}
  \sum\limits_{\Lambda,n,m} \hskip -.6 truecm \bigintsss\; T^a_{n\Lambda n}\Psi_{n,\Lambda,n}(\rho,\varphi_1,\varphi_2)& \ \ \mathrm{PB}\ ,\\[5pt]
  \sum\limits_{k,n,m} T^a_{nmk} \Phi_{nmk}(\rho,\varphi_1,\varphi_2)&\ \ \mathrm{LB} \ .
  \end{array}\right.
\eeqa
$\g(\SL)$ is also a Lie algebra (namely, the analogue of the loop  algebra in the case of $\SL$), and its Lie brackets can be obtained by means of the structure constant of $\g$ and the Clebsch-Gordan coefficients of $\SL$:
\beqa
\label{eq:loop}
  \begin{array}{llll}
    \big[ T^{a}_{m\Lambda n}, T^{a'}_{m'\Lambda'n'} \big]=if^{aa'}{}_{a''}
 \sum\limits_{\Lambda''} \hskip -.4 truecm {\bigintsss{}}\;
 c_{\Lambda \Lambda'}^{\Lambda''}{}_{mm'nn'}  T^{a''}_{m+m'\Lambda''n+n'}&\ \ \text{PB},\\
 \big[ T^{a}_{m nk}, T^{a'}_{m'n'k'} \big]=if^{aa'}{}_{a''}
 \sum\limits_{k''} c_{k k '}^{k''}{}_{mm'nn'}  T^{a''}_{m+m' n+n'k''}&\ \ \text{LB}.
 \end{array}\nn
  \eeqa
 Moreover, $\g(\SL)$ can be endowed with the scalar product
\beqa
\Big<X, Y \Big>_1= \frac 1 {4\pi^2}\int \limits_{0}^{+ \infty} \text{d} \rho \sinh \rho \cosh\rho
\int \limits_0^{2\pi} \text{d} \varphi_1 \int \limits_0^{2\pi} \text{d} \varphi_2 \Big<X,Y\Big>_0. \nn
\eeqa

(3): We introduce the maximal set of Hermitean commuting operators, namely $L_0,R_0$, with obvious commutation relations:
\beqa
\label{eq:LRT}
\begin{array}{lllllll}
\big[L_0,T^{a}_{n\Lambda m}\big]&=& n T^{a}_{n\Lambda m}\ ,& \big[R_0,T^{a}_{n\Lambda m}\big]&=& m T^{a}_{m\Lambda n}\ & \text{PB},\\
\big[L_0,T^{a}_{nmk}\big]&=& n T^{a}_{mnk}\ ,& \big[R_0,T^{a}_{mnk}\big]&=& m T^{a}_{mnk}\  &\text{LB}.
\end{array}
\eeqa

(4):  We introduce central extensions, by means of the two-cocycle
\beqa
\omega_\gamma(X,Y) = \frac 1{4\pi^2} \int \limits_{0}^{+ \infty} \text{d} \rho \sinh \rho \cosh\rho
\int \limits_0^{2\pi} \text{d} \varphi_1 \int \limits_0^{2\pi} \text{d} \varphi_2
\Big<X,\d Y\Big>_0\wedge \gamma \ ,\nn
\eeqa
where $\gamma$ is a closed two-form and $\d Y$ is the exterior derivative of $Y$. Strictly speaking, any closed two-form can be associated to a central extension. In particular, in duality with the Hermitean commuting operators $L_0$ and $R_0$, we respectively introduce the following two
central extensions  \cite{ram}:
\beqa
\omega_L(X,Y)&=&-\frac {k_L} {4\pi^2}  \int\limits _0^{+\infty} \d\rho \sinh \rho \cosh \rho \int \limits_0^{2\pi} \d \varphi_1  \int \limits_0^{2\pi}\d \varphi_2
\;\Big<X,L_0 Y\Big>_0,
\nn\\
\omega_R(X,Y)&=& -\frac {k_R} {4\pi^2}\int\limits _0^{+\infty} \d\rho \sinh \rho \cosh \rho \int \limits_0^{2\pi} \d \varphi_1  \int \limits_0^{2\pi}\d \varphi_2
\;\Big<X,R_0 Y\Big>_0\ . \nn
\eeqa
Confining ourselves to report only the expression of the central extension associated to $\omega_L$, and using the notation $\delta_{a+b}=\delta_{a,-b}$, we have
\beqa
\begin{array}{ll}
\left.
\begin{array}{lll}
\omega_L(T^{a}_{n\lambda\eta m}, T^{a' }_{n'\lambda '\eta'm'})&=&  n k_L g^{aa'} \delta_{\lambda,\lambda'} \delta_{\eta+\eta'} \delta_{m+m'} \delta_{n+n'}\\
\omega_L(T^{a}_{n i\sigma\epsilon m}, T^{a'}_{n' i\sigma'\epsilon'm'})&=&nk_Lg^{aa'} \frac{\delta(\sigma-\sigma')}{\sigma\tanh \pi(\sigma+i\epsilon)} \delta_{\epsilon,\epsilon'} \delta_{m+m'} \delta_{n+n'}
\end{array}\right\}& \text{PB},\nn\\ \\
\begin{array}{lll}
 \hskip .55truecm  \omega_L(T^{a}_{n\Lambda m}, T^{a'}_{n'\Lambda'm'}) &=& \;n k_L\;g^{aa'} \delta(\Lambda,\Lambda')  \delta_{m+m'} \delta_{n+n'}
\end{array}&\text{LB},
\end{array}
\eeqa
where we introduced
\beqa
\delta(\Lambda,\Lambda') =
\left\{
\begin{array}{cc}
 \delta_{\lambda,\lambda'} \delta_{\eta+\eta'} & \Lambda=(\lambda,\eta)\ ,\ \  \Lambda'=(\lambda',\eta')\\
 \frac{\delta(\sigma-\sigma')}{\sigma\tanh \pi(\sigma+i\epsilon)} \delta_{\epsilon,\epsilon'} &
 \Lambda=(i\sigma,\epsilon)\ , \ \  \Lambda'=(i\sigma', \epsilon')\\
 0&\text{elsewhere},
\end{array}
\right.\nn
\eeqa
such that $\delta(\Lambda,\Lambda')$ allows two write down the same formula also for the PB:
\beqa
\omega_L(T^{a}_{n\Lambda m}, T^{a'}_{n'\Lambda'm'}) &=& \;n k_L\;g^{aa'} \delta(\Lambda,\Lambda')  \delta_{m+m'} \delta_{n+n'}.\nn
\eeqa
\\
(5) : The Kac-Moody Lie algebra associated to $\SL$ is then defined as 
\beqa
\widehat{\g}(\SL):= \g(\SL) \cup\big\{L_0,R_0,k_L, k_R\big\}, \nn
\eeqa
and the points (1)-(4) above lead to the Lie brackets
\beqa
\label{eq:LieP}
\big[T^{a}_{m \Lambda n}, T^{a'}_{m'\Lambda'n'}\big] &=&i f^{a a'}{}_{a''} C_{\Lambda, \Lambda'}^{\Lambda''}{}_{m,m',n, n'} T^{a''}_{m+m'\Lambda''n+n'}\nn\\
&&\hskip 1.5truecm
+ (mk_L+nk_R) \delta(\Lambda,\Lambda') \delta_{m+m'} \delta_{n+n'}\ , \nn \\
\big[L_0,T^{a }_{m\Lambda n}\big]&=& m T^{a}_{m\Lambda n}\ , \\
 \big[R_0,T^{a,}_{m\Lambda n}\big]&=& n T^{a}_{m\Lambda n} \nn
\eeqa
in the PB, and
\beqa
\label{eq:LieL}
\big[T^{a}_{m n k}, T^{a'}_{m'n' k'}\big] &=&i f^{a a'}{}_{a''} C_{k k'}^{k''}{}_{m,m',n,n'} T^{a'' }_{m+m'n+n' k''}\nn\\
&&\hskip 1.5truecm
+ (mk_L+nk_R) \delta_{kk'} \delta_{m+m'} \delta_{n+n'}\ , \nn \\
\big[L_0,T^{a }_{mn  k}\big]&=& m T^{a}_{mn k}\ , \\
 \big[R_0,T^{a}_{mn k}\big]&=& n T^{a}_{mn k} \nn
\eeqa
in the LB. Since the LB can be expanded into the PB (see \eqref{eq:PL}) and conversely the PB can be expanded into
the LB (see \eqref{eq:LP}), it follows that the two presentations \eqref{eq:LieP} and \eqref{eq:LieL} of the algebra $\widehat{\g}(\SL)$ are equivalent.

It is also here worth remarking that the Lie brackets of $T^a(\rho,\varphi_1,\varphi_2)$'s can be written using the usual current algebra with a Schwinger term \cite{Scw},
\beqa
\big[T^a(\rho,\varphi_1,\varphi_2), T^{a'}(\rho',\varphi'_1,\varphi'_2)\big]&=&
\Big(if^{a a'}{}_{a''} T^{a''}(\rho',\varphi'_1,\varphi'_2) - g^{ab}(k_L L_0 +k_R R_0)\Big)\times \nn\\
&& \delta(\sinh^2 \rho-\sinh^2\rho') \delta(\varphi_1-\varphi'_1)  \delta(\varphi_2 -\varphi'_2),  \nn
\eeqa
which then reproduces either \eqref{eq:LieP} or \eqref{eq:LieL} \cite{ram}.

\subsection{Properties and Applications}
The Kac-Moody Lie algebra \eqref{eq:LieP} (or, equivalently, \eqref{eq:LieL}) has interesting properties, which we will briefly mention (again, for a more detailed discussion, see \cite{ram}).\\

Let us consider the root structure of $\widehat{\g}(\SL)$ in the LB. Assume that $\g$ is a rank$-\ell$ Lie algebra. Let $\Sigma$ be the roots of the Lie algebra $\g$ with respect to a Cartan subalgebra of $\g$.
Let  $\{H^i, i=1,\cdots,\ell\}$ be
the  Cartan subalgebra and let $E_\alpha, \alpha
\in \Sigma$  be the corresponding root vector.
The root space of $\widehat{\g}(\SL)$  then reduces to (with obvious notations; see \cite{ram} for more details concerning notations)
\beqa
\begin{split}
  \mathfrak{g}_{(\alpha,m,n)} &= \Big\{E_{\alpha,m n k} \ , k\in \mathbb N \Big\}\ , \ \ \alpha \in\Sigma, m,m \in \mathbb Z + \epsilon, \epsilon=0,\frac12,\\
  \mathfrak{g}_{(0,m,n)} &= \Big\{H^i_{mnk}\ , i=1,\cdots,\ell\ ,  k\in \mathbb N  \Big\}\ ,  m,n \in \mathbb Z+\epsilon, \epsilon=0,\frac12,
  \end{split}\nn
\eeqa
with the obvious commutation relations
\beqa
\Big[\mathfrak{g}_{(\alpha,m,n)},\mathfrak{g}_{(\beta,p,q)} \Big]&\subset &\mathfrak{g}_{(\alpha+\beta,m+p,n+q)}\ , \ \ \text{if} \ \ \alpha+\beta\in \Sigma \cup\{0\} \ \  \text{(and $=0$ otherwise),}\nn\\
  \Big[\mathfrak{g}_{(\alpha,m n)},\mathfrak{g}_{(0,pq)} \Big]&\subset& \mathfrak{g}_{(\alpha,m+p,n+q)}\ .\nn
  \eeqa
  In particular, it turns out that $\widehat{\g}(\SL)$ is an infinite rank Lie algebra \cite{ram}.\\

In \cite{rmm}, Kac-Moody algebras associated to a compact Lie group $G_c$ (or the coset $G_c/H$, with $H$ here denoting any proper subgroup of $G$) have been introduced for the first time. Such Lie algebras associated to $G_c$ and the Kac-Moody Lie algebras reported in this short note (and introduced in \cite{ram}) are of the same type, since they are constructed along the same principles, the only difference being the compactness or non-compactness of the underlying Lie group. In this respect, both such algebras are generalisations of affine Lie algebras, which are algebras associated to the Abelian Lie group $U(1)$.
On the other hand, it is well known that another different generalisation of affine Lie algebras exists, given by the Kac-Moody algebras defined through a generalised  Cartan matrix \cite{Kac,Moo}. It is interesting to observe that these two types of Kac-Moody algebras exhibit fundamentally different features. Indeed,  Kac-Moody algebras associated to $G_c$ or $\SL$ lack of a system of simple roots, but nevertheless all roots are explicitly known \cite{ram,rmm}, whereas for  Kac-Moody algebras associated to a (generalised) Cartan matrix
there always exists a system of simple roots, but their generators are only iteratively known in terms of the Serre--Chevalley relations.
\\

If $G$ is a non-compact Lie group and $H$ is its maximal compact subgroup, then the embedding of $H$ in $G$ is always symmetric, and the harmonic analysis on the coset $G/H$ is known. In \cite{ram} we also defined a Kac-Moody algebra associated to the coset $\SL/U(1)$, again using a twofold approach, namely by exploiting the Plancherel Theorem or by identifying a Hilbert basis of $\SL/U(1)$. Such  a space naturally appears in physical theories, such as supergravity, as the target space of scalar fields, namely in the simplest case in which only one (complex%
\footnote{%
The complex nature of the scalar field is imposed by ($\mathcal{N}=1$ or $%
\mathcal{N}=2$-extended) supersymmetry (see below).}) scalar field occurs,
and it coordinatises a scalar manifold which is (locally) isomorphic to the K%
\"{a}hler symmetric coset\footnote{%
The notation \textquotedblleft $\simeq $\textquotedblright\ here denotes local
isomorphism of symmetric spaces.}%
\beqa
\frac{SU(1,1)}{U(1)}\simeq \frac{SL(2,\mathbb{R})}{U(1)}\simeq \frac{Sp(2,%
\mathbb{R})}{U(1)}.  \label{m}
\eeqa
In presence of $\mathcal{N}\geqslant 2$ supercharges, the non-compact,
Riemannian, locally symmetric, constant curvature, rank-1 space (\ref{m})
occurs three times as scalar manifold of a supergravity theory in $D=3+1$.
 These theories are : i) \textquotedblleft pure\textquotedblright $\mathcal{N}=4$ supergravity
(only containing the gravity multiplet); ii) $\mathcal{N}=2$ supergravity \textit{minimally} coupled to $1$ vector
multiplet; iii) $\mathcal{N}=2$ supergravity obtained as dimensional reduction from $%
\mathcal{N}=2$ \textquotedblleft pure\textquotedblright\ (minimal)
supergravity in $D=5$ (the so-called $T^{3}$ model, with $1$ vector
multiplet). For further discussion, we address the reader to \cite{ram}).

\section{Conclusions}
We have reported on some results recently obtained in \cite{ram}, in which a novel type of infinite-dimensional, Kac-Moody algebra, related to the non-compact (split, rank-1) Lie group $SL(2,\mathbb{R})$, was introduced. These results are non-trivial generalisations of analogous findings obtained for the first time in \cite{rmm} for compact Lie groups, since harmonic analysis on non-compact manifolds is in general much more involved than its compact counterpart. We strove to present the two (different but equivalent) faces of our investigation : on the one hand,  the Plancherel
Theorem has been used, while on the other hand we have determined a Hilbert basis on the space
of square-integrable functions $L^{2}\left( SL(2,\mathbb{R})\right)$.

The appearance of $SL(2,\mathbb{R})/U(1)$ as a scalar manifold of models of Maxwell-Einstein supergravity in $D=3+1$ space-time dimensions suggests the application of $\widehat{\mathfrak{g}}\left( SL(2,\mathbb{R})/U(1)\right) $ in the context of supergravity. Since, as mentioned, the class of Kac-Moody algebras introduced in \cite{ram} (and reviewed here) exhibits strikingly different features than the usual Kac-Moody extensions of (semi)simple Lie algebras (which have found extensive applications in supergravity, see e.g. \cite{West, Nicolai, West-VP-Riccioni}), this possibility would pave the way to novel applications and results in the investigation of (possibly effective) theories of quantum gravity. The infinite rank of $\widehat{\mathfrak{g}}\left( SL(2,\mathbb{R})/U(1)\right) $ may constitute a technical issue, but we hope to report on this intriguing venue of research in future work.

\section*{Acknowledgments}
We take again the chance to gratefully thank Viktor Losert for having provided us with an Hilbert basis of $\LL$. The work of AM is supported by a “Maria Zambrano” distinguished researcher fellowship at the University of Murcia, Spain, ﬁnanced by the European Union within the NextGenerationEU programme. The work of RCS has been supported by the grant PID2023-148373NB-I00 funded by MCIN /AEI /10.13039/501100011033 / FEDER, UE. 


\end{document}